\documentclass[sigconf,authorversion]{acmart}

\usepackage{enumitem}
\usepackage{tcolorbox}
\usepackage{listings}
\usepackage{booktabs}
\usepackage{longtable}
\usepackage{cleveref}
\usepackage{multirow}
\usepackage{siunitx}
\usepackage{colortbl}
\usepackage{caption}
\usepackage{makecell}
\usepackage{algorithm}
\usepackage{algpseudocode}
\usepackage{geometry}
\usepackage{pgfplots}
\usepackage{pgfplotstable}
\usepackage{xcolor}
\usepackage{graphicx}
\usepackage{adjustbox}
\usepackage{tikz}
\pgfplotsset{compat=1.18}
\usetikzlibrary{patterns}

\definecolor{darkGreen}{rgb}{0.0, 0.6, 0.0} 
\definecolor{darkRed}{rgb}{0.6, 0.0, 0.0} 
\colorlet{lightergray}{gray!5}

\tcbset{arc=2pt, boxrule=\heavyrulewidth, colback=lightergray, colframe=black}

\newcommand{\header}[1]{\subsubsection*{\emph{\textbf{#1}}}}

\newcommand{\rqi}{How effective are LLMs in resolving web performance issues through automated DOM-level modifications?}
\newcommand{\rqii}{What is the nature of changes made by LLMs for automated DOM-Level resolution of web performance issues?}

\begin{document}

\settopmatter{printacmref=false}
\renewcommand\footnotetextcopyrightpermission[1]{}
\setcopyright{none}
\pagestyle{plain}

\title{Evaluating the Use of LLMs for Automated DOM-Level Resolution of Web Performance Issues}

\author{Gideon Peters}
\affiliation{%
    \institution{Concordia University}
    \city{Montreal}
    \country{Canada}
}
\email{gi_peter@mail.concordia.ca}

\author{SayedHassan Khatoonabadi}
\affiliation{%
    \institution{Concordia University}
    \city{Montreal}
    \country{Canada}
}
\email{sayedhassan.khatoonabadi@concordia.ca}

\author{Emad Shihab}
\affiliation{%
    \institution{Concordia University}
    \city{Montreal}
    \country{Canada}
}
\email{emad.shihab@concordia.ca}

\renewcommand{\shortauthors}{Peters et al.}

\begin{abstract}
    Users demand fast, seamless webpage experiences, yet developers often struggle to meet these expectations within tight constraints. Performance optimization, while critical, is a time-consuming and often manual process. One of the most complex tasks in this domain is modifying the Document Object Model (DOM), which is why this study focuses on it. Recent advances in Large Language Models (LLMs) offer a promising avenue to automate this complex task, potentially transforming how developers address web performance issues. 
    This study evaluates the effectiveness of nine state-of-the-art LLMs for automated web performance issue resolution.
    For this purpose, we first extracted the DOM trees of 15 popular webpages (e.g., Facebook), and then we used Lighthouse to retrieve their performance audit reports. Subsequently, we passed the extracted DOM trees and corresponding audits to each model for resolution. 
    Our study considers 7 unique audit categories, revealing that LLMs universally excel at SEO \& Accessibility issues. However, their efficacy in performance-critical DOM manipulations is mixed. While high-performing models like GPT-4.1 delivered significant reductions in areas like \textit{Initial Load}, \textit{Interactivity}, and \textit{Network Optimization} (e.g., 46.52\% to 48.68\% audit incidence reductions), others, such as GPT-4o-mini, notably underperformed, consistently. A further analysis of these modifications showed a predominant additive strategy and frequent positional changes, alongside regressions particularly impacting \textit{Visual Stability}. Our findings define safe areas for automation (e.g., SEO and accessibility) and reveal the limits of DOM-level resolution, underscoring the need for hybrid, validated workflows. However, it critically underscores the need for careful model selection, understanding their specific modification patterns, and robust human oversight to ensure reliable web performance improvements.
    
\end{abstract}



\maketitle

\section{Introduction}
Web applications have become one of the primary ways users consume content on the internet~\citep{bocchi2016measuring}. Therefore, the importance of performant web applications cannot be overemphasized~\citep{towardsImprovingAccessibility}. A webpage's performance is a core non-functional requirement, as it impacts the overall user experience, engagement, and conversion ratios~\citep{enterpriseWebOptimisation, indonesiaLighthouse}. As such, web performance engineering remains an unnegotiable component of the web development process. It requires a deep understanding of both the browser engine and application use cases~\citep{industrialWebPerformance}. Performance optimization involves various considerations including hardware (CPU and memory usage), server (API response times), and client-side factors (DOM size, image optimizations, on-demand loading, and omni-channel experience)~\citep{persson2020javascript, chkec2019performance}.

This study focuses on the client-side, specifically the Document Object Model (DOM), which is central to how browsers interpret, render, and interact with webpages~\citep{wood1998document, dompletion2014}. The DOM also significantly impacts hardware and server performance---complex DOM structures increase CPU and memory usage, slowing performance---especially on resource-limited devices. Additionally, large DOM payloads can strain server response times~\citep{persson2020javascript, chkec2019performance, goel2020web}. Optimizing the DOM is challenging~\citep{subraya2000object}, requiring detailed analysis and targeted modifications to balance functionality and performance. Traditionally, addressing DOM inefficiencies has relied on manual interventions and automated tools with limited scope~\citep{macakouglu2022web}. However, the growing complexity of web applications demands more sophisticated solutions~\citep{bocchi2016measuring}. 

Large Language Models (LLMs) present a promising approach to address these challenges. They have transformed numerous software engineering tasks by leveraging their ability to understand and generate human-like text~\citep{hou2023large, seAdapt2024, hadi2024large}. Trained on massive corpora, including HTML documents from public repositories, LLMs are uniquely positioned to tackle challenges in web development~\citep{endUserWebsiteGeneration2023, deng2024mind2web}. Their applications extend beyond code generation to include tasks like web security~\citep{toth2024llms}, automated testing~\citep{nass2023improving}, and accessibility improvements~\citep{lopez2024turning}. However, the effectiveness of LLMs for web performance optimization, particularly in modifying the DOM to address performance issues, has not yet been systematically explored.

To fill this knowledge gap, we aim to explore the usefulness and challenges of using LLMs for automating web performance resolutions. For this purpose, we extract the DOM trees of 15 popular webpages, and we generate audit reports for these extracted DOM trees. We then assess the effectiveness of nine state-of-the-art LLMs, including GPT-4.1, Claude 3.7 Sonnet, DeepSeek R1 \& V3, and GPT-4o-mini—to resolve these audits by passing the audit along with the DOM tree to the model. Finally, we generate new audit reports for the modified DOM trees and compare the incidence of the initial audits before and after modification. In summary, we aim to answer the following research questions:

\begin{itemize}
    \item[\textbf{RQ1:}] \textbf{\rqi} Performance optimization can be tedious, requiring web developers to run performance tests, and implement required fixes~\citep{herivcko2021towards, mcgill2023towards}.
    We explore the ability of LLMs to resolve performance issues identified by Lighthouse audits across 15 webpages. Our findings indicate that LLMs achieved a 100\% reduction in SEO \& Accessibility issues.
    However, for performance-critical issues, effectiveness was mixed and highly model-dependent, with some models showing significant gains while others notably introduced regressions, particularly impacting visual stability.
    
    \item[\textbf{RQ2:}] \textbf{\rqii} To understand how LLMs modify the DOM, we analyzed differences between the original and LLM-modified HTML pages for nine state-of-the-art models across 15 webpages. We identified modifications including element and attribute additions, removals, type changes, and positional shifts. Most LLMs used a predominantly additive strategy, with GPT-4o-mini uniquely removing more elements than it added. Frequent positional changes also occurred, typically at shallower DOM depths.
\end{itemize}

Our study offers insights for web developers, LLM providers, and researchers on automating web performance remediation. We focus on DOM-level edits to compare models under identical, language-agnostic inputs and maintain reproducibility across sites. This baseline is essential since every web stack ultimately renders to the DOM~\citep{googlewebdom}. However, performance regressions can also stem from the interaction of DOM, CSS, and JavaScript; thus, our findings should be applied conservatively to SPA or dynamic contexts and verified through CI/CD checks. Focusing on DOM-level changes enables direct, interpretable fixes that integrate seamlessly into existing workflows while avoiding build- or server-side confounds~\citep{nass2023improving}.

\header{Our Contributions.} In summary, we make the following contributions in this paper:

\begin{itemize}[leftmargin=*]
    \item We conducted extensive experiments using nine LLMs on DOM trees from 15 popular webpages, providing a comprehensive evaluation of their effectiveness in automated web performance issue resolution.
    \item We provide a token-aware chunking strategy for DOM trees based on a predefined token threshold to enable processing by LLMs for the task of web performance issue resolution.
    \item We identified seven distinct audit categories, and provide a detailed quantitative analysis of LLM changes implemented with respect to these audits.
    \item We synthesize actionable insights and implications for web developers, LLM providers, and the research community, guiding future development towards more robust and reliable AI-driven web performance optimization.
    \item To promote the reproducibility of our study and facilitate future research on this topic, we publicly share our scripts and dataset online~\citep{replicationPackage}.
\end{itemize}

\header{Paper Organization.} The rest of the paper is organized as follows. \Cref{sec:background} outlines the necessary background for our study. \Cref{sec:study_design} details the study design. \Cref{sec:rqi,sec:rqii} examine the effectiveness of LLMs in resolving web performance issues in the DOM and the nature of changes implemented. \Cref{sec:discussion} discusses our findings. \Cref{sec:limitations} outlines the study limitations. \Cref{sec:related_work} reviews related literature and \Cref{sec:conclusion} concludes the paper.


\section{Background}
\label{sec:background}
This section introduces key concepts and technologies relevant to our study. 

\subsection{DOM Trees}
\begin{figure}
    \centering
    \small
    \includegraphics[width=.7\linewidth]{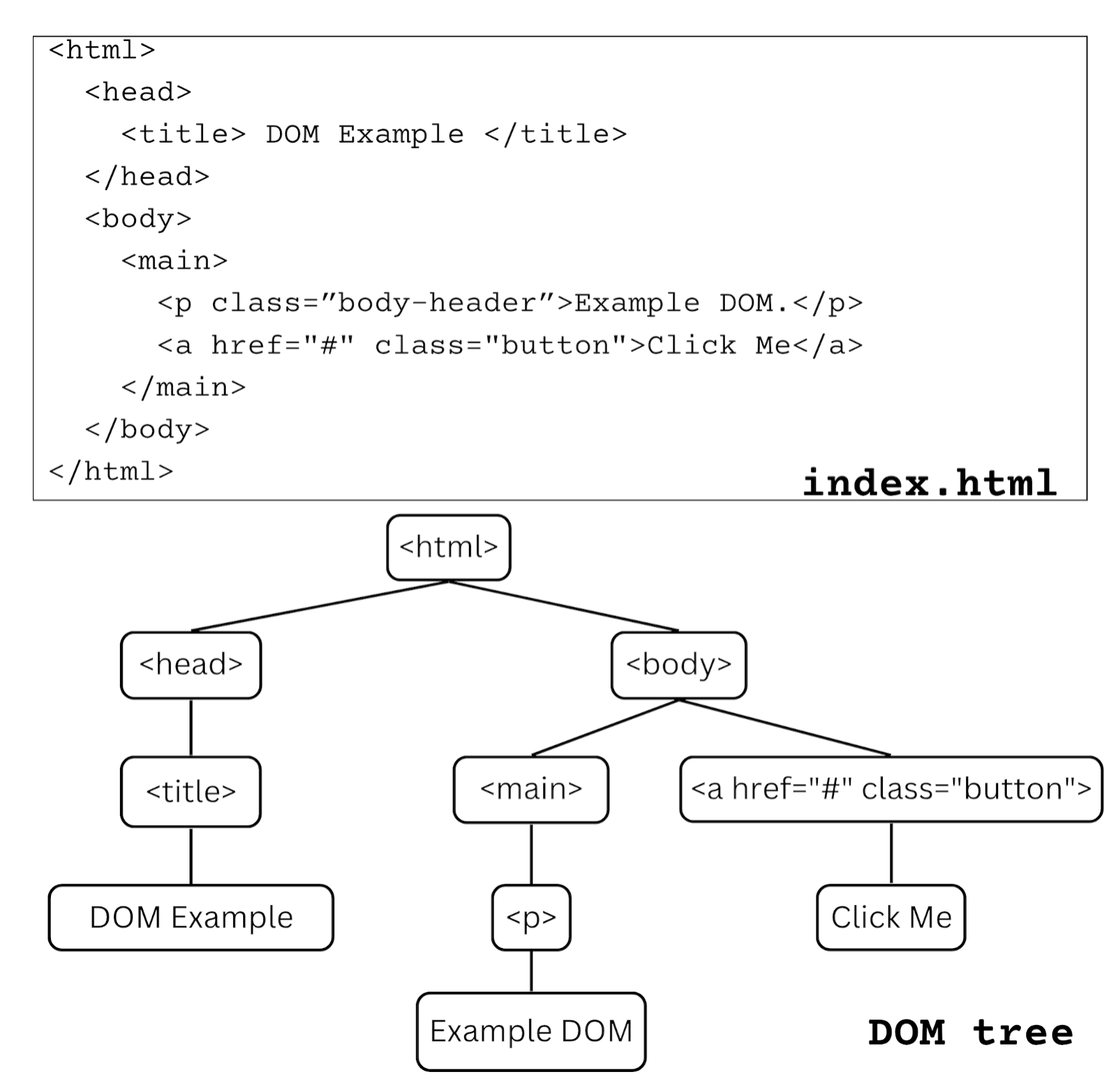}
    \caption{Example structure of a DOM tree}
    \label{fig:dom-tree-example}
    \Description{DOM tree structure}
\end{figure}

The Document Object Model (DOM) represents a webpage's structure and content as a tree-like hierarchy of nodes, where each node corresponds to an HTML element, attribute, or text~\citep{wood2000document}. This hierarchy enables programmatic access and manipulation of webpage elements. The DOM is fundamental to web development, allowing dynamic updates and interaction with web content. Through DOM manipulation, developers can: (i) dynamically alter webpage structure, style, and content, and (ii) respond to user interactions. \Cref{fig:dom-tree-example} illustrates a DOM tree, showing its HTML code representation on the top and its hierarchical structure on the bottom.

It originates with the \texttt{<html>} element as its \textbf{root node}, which serves as the parent to both the \texttt{<head>} and \texttt{<body>} elements. This hierarchy extends through \textbf{parent-child relationships}, where elements are nested within one another (e.g., \texttt{<body>} typically contains \texttt{<header>}, \texttt{<main>}, and \texttt{<footer>}). Additionally, nodes at the same hierarchical level exhibit \textbf{sibling relationships}, such as \texttt{<p>} and \texttt{<a>} elements found within a \texttt{<main>} section.

Furthermore, DOM trees contain various element types: \textbf{Text} for content (e.g., \textit{Example DOM} in \texttt{<p>Example DOM</p>}); \textbf{Comment} for unrendered documentation (e.g., \texttt{{{<!-- -->}}}); \textbf{Tag} as core HTML elements (e.g., \texttt{<p>}) with optional attributes (e.g., \textit{class} in \texttt{<a class="button">Link</a>}); \textbf{Script} for embedding or referencing JavaScript (e.g., \texttt{<script src="app.js"></script>}); and \textbf{Stylesheet} for CSS rules (e.g., \texttt{<style>p \{ color: red; \}</style>}).


These element types are combined in various ways and for different purposes across webpages~\citep{world2004document}. In performance optimization, DOM tree size and complexity significantly impact webpage load time and responsiveness, as larger and more deeply nested trees demand greater computational resources for rendering and processing~\citep{gizas2015performance, kuparinen2023improving}.

\subsection{Performance Audits}
Performance audits systematically evaluate webpages to assess their performance, identify bottlenecks, and recommend improvements~\citep{shivakumar2020modern}. These audits aim to ensure sites meet performance goals, such as fast load times, smooth interactions, and efficient resource usage. These are critical for user experience, search engine rankings, and business outcomes~\citep{herivcko2021towards}. 

Our study utilizes \texttt{Lighthouse}, an open-source Chromium-based tool developed by Google for the measurement and improvement of webpage performance~\citep{lighthouseOverview, scherer2020hands, psInsights}. We selected \texttt{Lighthouse} due to its extensibility and widespread community adoption~\citep{herivcko2021towards, towardsImprovingAccessibility, dinizWebPerformanceEval2022}. It analyzes webpages, generating actionable reports on performance, accessibility, SEO, and progressive web apps. 
\texttt{Lighthouse} offers numerous configuration flags allowing tailored audits for specific use cases, environments, or requirements.
\texttt{Lighthouse} audit reports include key performance metrics like First Contentful Paint (FCP), Largest Contentful Paint (LCP), and Cumulative Layout Shift (CLS)~\citep{understandingLighthouseMetrics, towardsImprovingAccessibility, herivcko2021towards}. Each audit highlights an issue or suggestion based on standard practices and is keyed by its name and contains the following properties:
\begin{itemize}[leftmargin=*]
    \item \textbf{id}: A unique identifier identical to the audit key.
    \item \textbf{title}: A brief summary of the audit.
    \item \textbf{description}: A detailed explanation of what the audit assesses and its significance.
    \item \textbf{score}: A numeric or categorical value indicating the audit's result.
    \item \textbf{scoreDisplayMode}: Denotes how the score is interpreted, with values; \texttt{informative}, \texttt{notApplicable}, \texttt{manual}, or \texttt{error}).
    \item \textbf{displayValue}: A contextual measurement supplementing the score.
    \item \textbf{details}: Provided when an audit fails, offering insight into the issue and potential resolutions. This may include responsible element types, value headings, or affected items like specific DOM elements, location parameters, resource URLs, or data points. 
\end{itemize}

\section{Study Design}
\label{sec:study_design}
In this section, we describe our dataset, the performance audits, the environment configurations, considerations for our LLM selection, the chunking strategy, and the evaluation metric used in this study.

\subsection{Dataset}
To conduct our study, we first select 15 real-world webpages at random from the Alexa Top 500 list~\citep{alexawebsites}, which features top-ranked webpages on the web. We chose this list due to its popularity and prior use in research~\citep{chan2020investigating, mahajan2018automated, ocariza2011javascript}. Each webpage selected is a homepage, the main entry point for users. Since homepages typically receive the highest traffic~\citep{geissler2006influence, papacharissi2002self, singh2005understanding}, optimizing their performance is particularly relevant to our study.

Our dataset comprises webpages from four different categories: \textit{Shopping webpages (4)}, \textit{Professional webpages (2)}, \textit{Social webpages (6)}, and \textit{Entertainment webpages (3)}. The full breakdown can be found in our replication package~\citep{replicationPackage}.
\Cref{tab:dataset_descriptive_stats} presents descriptive statistics for the webpages in our dataset. DOM Tree Depth, the maximum depth of nested HTML elements, ranging from 4 to 32, indicating diverse structural complexity. The number of chunks (\# chunks), varying from 2 to 17, reflecting varied content modularity across webpages. 
\emph{Total Audits} averaged 25.5 from 18 to 39, providing substantial per-page data. Finally, \emph{LHS} (Lighthouse Score) averaged 44.7\%, ranging from 12\% to 90\%, highlighting significant variability in webpage performance.

\begin{table}[h]
\caption{Descriptive Statistics of the Webpages in Our Dataset}
\centering
\small
\begin{tabular}{@{}lrrr@{}}
\toprule
\textbf{Statistic} & \textbf{Mean} & \textbf{Minimum} & \textbf{Maximum} \\
\midrule
DOM Tree Depth & 18 & 4 & 32 \\
\# Chunks & 5.2 & 2 & 17 \\
Total Audits & 25.5 & 18 & 39 \\
LHS (\%) & 44.7 & 12 & 90 \\
\bottomrule
\end{tabular}
\label{tab:dataset_descriptive_stats}
\end{table}

\begin{figure}
    \centering
    \includegraphics[width=\linewidth]{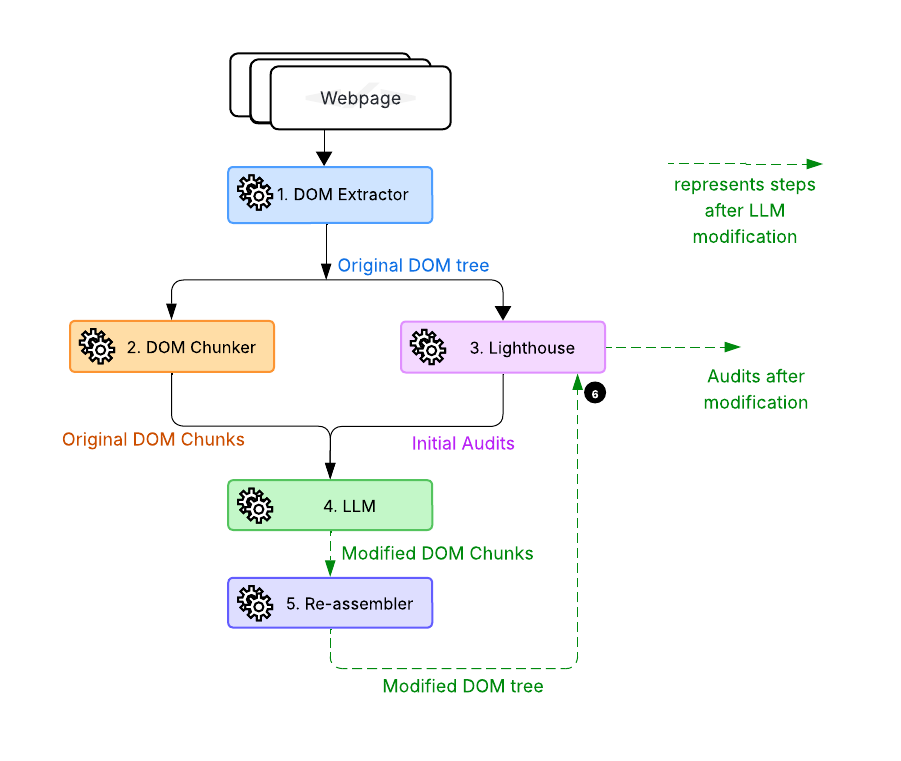}
    \caption{Overview of our experiment workflow}
    \label{fig:study_design}
\end{figure}

\Cref{fig:study_design} shows the entire workflow for our experiments. For each webpage, it comprises the following six main stages:

\begin{enumerate}[leftmargin=*]
    \item \textbf{DOM Extraction}: We begin by extracting \textit{original DOM trees} from the webpage. Python's \texttt{requests} package was used to fetch webpages, and their DOM trees were then extracted and parsed with \texttt{BeautifulSoup}~\citep{pybsoup}.
    \item \textbf{DOM Chunking}: To accommodate the LLMs' context window and output token limitations, we split the DOM tree into smaller chunks to obtain the \textit{original DOM chunks}.
    \item \textbf{Initial Audit Report Generation}: The extracted DOM tree is then passed to \texttt{Lighthouse} to generate \textit{initial audit reports}, which establish a benchmark for issues to be resolved by the LLMs.
    \item \textbf{LLM Modification}: Each chunk is then provided to the LLM, along with the corresponding audit reports, instructing it to make modifications to resolve the identified issues. This process is applied to every chunk with the LLM returning the \textit{modified DOM chunk} in every iteration.
    \item \textbf{Re-assembly}: After processing all original chunks, the modified chunks are reassembled into a complete \textit{modified DOM tree}.
    \item \textbf{Post-Modification Audit Report Generation}: A subsequent audit report is generated from this reassembled tree to capture the \textit{audits after LLM modification}. This allows for a quantitative comparison between the initial audit reports and those obtained from the modified DOM trees to assess the LLM's effectiveness.
\end{enumerate}

The subsequent sections provide a detailed explanation of this workflow and our considerations.

\subsection{LLM Selection}
To evaluate the potential of LLMs in automating web performance issue resolution, we selected a diverse set of state-of-the-art LLMs that vary in reasoning capability, architecture, context window limits, max output tokens, and model size. Table~\ref{tab:llm-specs-params} provides a detailed list of the models and their specifications. Our goal is to compare how different LLMs process DOM trees in conjunction with performance audits, and to understand how model characteristics influence the types and quality of generated modifications, thereby enhancing the generalizability of our findings. All LLMs used zero-shot prompting with the same prompt structure (provided in the replication package). We fixed temperature = 0.0 to minimize variance~\citep{doderlein2025piloting}. Max output tokens and chunk sizes followed the smallest-window model (GPT-4o-mini) to ensure fairness across models.
\begin{table}
\caption{LLM Models Evaluated and Their Specifications}
\label{tab:llm-specs-params}
\centering
\begin{adjustbox}{max width=\linewidth}
\begin{tabular}{@{}lcccc@{}}
\toprule
\textbf{Model} & \textbf{Reasoning} & \textbf{Max O.T.} & \textbf{Context Window} & \textbf{Size} \\
\midrule
Claude 3.7 Sonnet (R) & Yes & 128K & 200K & -- \\
Claude 3.7 Sonnet & No & 128K & 200K & -- \\
DeepSeek V3 & No & 32K & 131K & 671B \\
DeepSeek R1 & Yes & 32K & 128K & 685B \\
Llama3.3 70B & No & 40K & 128K & 70B \\
GPT-4.1 & No & 32K & 1M+ & -- \\
o4-mini & Yes & 100K & 200K & -- \\
GPT-4o-mini & No & 16K & 128K & -- \\
Qwen2.5 32B-Instruct & No & 128K & 131K & 32B \\
\bottomrule
\end{tabular}
\end{adjustbox}
\end{table}

\subsection{DOM Chunking}
To accommodate the varying output token limits of the LLMs in our evaluation, we employ a conservative chunking strategy. This approach is based on the model with the smallest maximum output size, specifically GPT-4o-mini, which has an output token limit of 16K~\citep{gpt4oInfo}. This ensures all models can process identical DOM chunks for a consistent and fair comparison.

We implement this by limiting DOM chunks to 15K tokens, reserving 1K tokens for LLM-induced modifications. Token counts are estimated using OpenAI's \texttt{Tiktoken} package~\citep{tiktoken}. Chunks exceeding 15K tokens are recursively split to ensure all webpage elements are assessed.

This strategy prevents token truncation, incomplete responses, resource inefficiencies during inference, and error propagation in downstream evaluations. Our approach traverses the DOM tree in a depth-first search, grouping nodes into chunks that never exceed the 15K threshold. This is done with attention to semantic structure and element types; for example, specific preservation strategies were applied to certain element types:

\begin{itemize}[leftmargin=*]
    \item \textbf{Text Nodes:} We preserved these text elements as-is, explicitly excluding them from any splitting or chunking operations to maintain the integrity of inline text content.
    \item \textbf{Comment:} HTML comments were left untouched to avoid losing useful annotations or developer metadata. We applied no chunking or transformations to these nodes.
    \item \textbf{Tag:} Before splitting, we stored all initial tag attributes for comparison and use during reassembly, ensuring tags and their associated attributes remained intact. The split was then performed recursively, accurately representing every element and its descendants. Each chunk was uniquely identified by a UUID to ensure accurate reassembly.
    \item \textbf{Script:} We also stored script elements before chunking, reincorporating them during HTML reassembly. This preserved the logic and interactivity defined by scripts.
    \item \textbf{Stylesheet:} Similarly, style rules were stored before chunking and merged back during reassembly. Preserving these styles maintains the visual fidelity of the webpage.
\end{itemize}

To validate the integrity of our chunking strategy, we reassembled all chunks before any modifications, confirming that the structure and content remained unchanged. To ensure there are no unintended alterations, we checked a popular metric known as \textbf{Tree Edit Distance} to quantify the difference between the original and reassembled DOMs~\cite{shin2021learning, leithner2018domdiff}.
For all webpages processed, we observed a \textbf{Tree Edit Distance of 0}. This indicates that the reassembled DOMs were identical to the original DOMs, with no structural or content alterations.

\subsection{Performance Audits}
To create an initial benchmark for what issues we attempt to resolve, we generate Lighthouse audit reports for the DOM trees before they are modified by the LLMs. We also generate audit reports for the LLM-modified DOM trees. These are used in our quantitative analysis. For the audit generation process, we make use of the following Lighthouse configuration flags:

\begin{itemize}[leftmargin=*]
    \item \textbf{headless}: Allowing Chrome to operate without a Graphical User Interface (GUI) and ensuring lower consumption of CPU and memory resources. Consequently, it facilitates the automation of the Lighthouse analysis limiting any interactions with the webpages during the process.
    \item \textbf{no-sandbox}: This disables the sandbox feature in Chrome which isolates web content and process. This is useful for our study as it bypasses security restrictions that may come up in our environment that could affect how pages are rendered.
    \item \textbf{disable-gpu}: Forcing the browser to render pages using the CPU instead of the GPU. This is to ensure consistency of results based on the allocated CPU resource in our environment.
\end{itemize}

From the audit reports generated, we exclude audits with \textit{scoreDisplayMode} values of \texttt{notApplicable}, \texttt{manual}, and \texttt{informative}, or \texttt{binary} scores of \texttt{1} as it indicates a pass, and these audits do not require any resolution in the DOM~\citep{understandingLighthouseMetrics}, unlike a binary score of 0, which indicates a needed resolution. This resulted in \textit{67 unique audits}, each of these audits as well as their descriptions can be found in our replication package~\citep{replicationPackage}. All of these audits were manually analyzed, and a classification was agreed upon by two authors, and any conflicts were resolved by the third author. This resulted in the establishment of seven audit categories. The categories are as follows:

\begin{itemize}[leftmargin=*]
    \item \textbf{Initial Load Performance:} Describes how quickly a page's essential content loads, e.g., \textit{"First Contentful Paint"} measures the time it takes for the first text or image to appear on the screen.
    
    \item \textbf{Interactivity Performance:} Focuses on how responsive the page is to user interactions, e.g., \textit{"Time to Interactive"} measures when the page becomes fully interactive, indicating when a user can reliably interact with the page.
    
    \item \textbf{Runtime Performance:} Assesses how efficiently JavaScript and other resources are executed during runtime. e.g., \textit{"JavaScript Execution Time"} measures the duration of JavaScript operations and their impact on page speed. 
    
    \item \textbf{Resource Optimization:} Evaluates the effectiveness of resource usage such as scripts, images, and stylesheets, e.g., \textit{"Unminified JavaScript"} flags large, uncompressed JavaScript files that could be optimized to reduce their size and improve performance. 
    
    \item \textbf{Network Optimization:} Measures the efficiency of network requests, including the number of requests and their size, e.g., \textit{"Reduce Server Response Time"} focuses on reducing latency and optimizing server performance to decrease load time. 
    
    \item \textbf{Visual Stability:} Focuses on preventing unexpected layout shifts during page load, e.g., \textit{"Cumulative Layout Shift"} tracks the unexpected shifting of elements as the page loads, impacting the user experience. 
    
    \item \textbf{SEO \& Accessibility:} Covers audits related to SEO and accessibility, e.g., \textit{"Accessibility Improvements"} flags issues that affect the usability of the website for users with disabilities, such as missing aria labels. These audits are relevant from a semantic point of view as opposed to the hierarchical context of the DOM~\citep{salem2025enhancing}.
\end{itemize}

In this paper, we use \textit{performance-critical} to mean audits that directly affect user-perceived latency, responsiveness, runtime efficiency, network/asset delivery, and layout stability. Concretely, this includes Initial Load Performance, Interactivity Performance, Runtime Performance, Resource Optimization, Network Optimization, and Visual Stability. By contrast, SEO \& Accessibility are treated as semantic audits rather than performance-critical.

\subsection{Environment Configurations}
To ensure the reproducibility and consistency of our performance audits, we utilized a Docker-isolated environment~\citep{merkel2014docker}. This approach mitigates variability stemming from diverse hardware configurations (CPU, RAM, GPU, etc.) by providing a standardized execution context, a common practice in web development for consistent builds.

Our Docker environment was hosted on a MacBook Pro 2018 featuring a 2.3GHz Quad-Core Intel Core i5 processor and 16GB of RAM. Key software versions used include Docker v27.0.3, Node v21.5.0, Lighthouse 12.2.0, and the python:3-9-slim Docker image runtime. The Docker container was allocated 1GB of shared memory to ensure sufficient resources for the Chromium browser used by Lighthouse.

\subsection{Benchmarking \& Evaluation Metric}
To calculate the distribution of performance issues across our dataset, we introduce the derived \textit{audit incidence ratio (AIR)} metric. It is a practical adaptation of reporting practices in tools like Lighthouse, which summarize how often specific audits are detected across sites to inform optimization priorities~\citep{lighthouseOverview}. The \textit{AIR} of an audit provides a quantitative measure of the extent to which it is observed in the dataset. A higher ratio indicates that the audit is more prevalent and affects a larger portion of the webpages, suggesting it could be a critical area to address for overall performance improvement. In contrast, a lower \textit{AIR} suggests that the issue affects fewer webpages. We define it as follows:
\begin{equation}
\text{AIR} = \frac{W_a}{W_{\text{total}}}
\end{equation}
where \(W_a\) is the number of unique webpages containing audit \(a\), and \(W_{\mathrm{total}}\) is the total number of webpages in the dataset.

To benchmark our approach, we compare the original \textit{AIR} for the extracted DOM trees with the \textit{AIRs} observed after applying modifications implemented by the LLMs. We call this comparison the percentage change in \textit{AIR}, calculated as follows:
\begin{equation}
\text{\% change in AIR} = 
\frac{M - O}{O} \times 100
\end{equation}

\section{RQ1: \rqi}
\label{sec:rqi}
\header{Motivation.} 
The iterative and time-consuming nature of web performance optimization presents a significant challenge for developers and negatively impacts user experience when performance is poor~\citep{herivcko2021towards, ocariza2011javascript, bocchi2016measuring, crescenzi2016impacts, basalla2021latency}. Investigating how LLMs emergent capabilities could automate the resolution of these issues offers a promising avenue for significant advancement in web development---our work specifically validates their ability to resolve web performance issues by making necessary changes to the DOM. We study LLM-driven DOM-level resolution to web performance issues, isolating this layer to avoid confounding build or server-side effects.

\header{Approach.} To assess the effectiveness of LLMs in resolving web performance issues in the DOM, we utilized the audits generated for the originally extracted webpages (see~\Cref{sec:study_design}). Subsequently, each webpage is split into chunks to address LLMs' output token limitation, as detailed in our study design (\Cref{sec:study_design}).
For each webpage, we then iteratively pass each chunk and the performance audits to the LLMs. Through zero-shot prompting~\citep{kojima2022large}, we instruct the LLMs to make the necessary changes to resolve contributing factors to these issues. Our prompt includes the audit key, title, description, and details (if any). Our replication package contains the full prompt structure utilized~\citep{replicationPackage}. 

Our prompt design specifically tailored the input to guide the LLMs in understanding complex DOM structures and performance audit requirements, considering their unique processing characteristics. It incorporates the following considerations:
\begin{itemize}[leftmargin=*]
    \item To address LLMs' inherent output token limitations, we designed our input strategy to feed the DOM in chunks, ensuring the LLM understood the incremental nature of the content and avoided changes that could disrupt the hierarchical DOM structure.
    \item We explicitly requested the LLM to specify modified sections and describe the changes, a crucial step for validating LLM-generated modifications and facilitating easy identification of affected areas.
    \item We provide some context to the LLM about the possibility that the DOM tree being processed is likely to be minified, uglified, or compressed as it is from a production website~\citep{vazquez2019slimming}. This is important as it lets the LLM know that some styles or scripts are already processed, hence further similar processing should be avoided to preserve the functionality of the webpage.
    \item We explicitly instruct the LLM to avoid any changes to the order, styles, and functionalities of the scripts present. This is done to preserve the core functionalities of the webpage.
    \item We constrain the LLMs to use the right formatting of modification comments in the respective sections, e.g., the HTML comment formatting for regular HTML elements and the style comment formatting for style scripts. This is done to avoid any parsing or build issues when the DOM tree is reassembled.
\end{itemize}

All modified chunks are then reassembled and a final audit report is generated on the updated webpage. This step helps to determine if the issues identified in the initial audit report have been resolved. To present this clearly, we conducted a quantitative analysis, comparing the audit reports of the original webpage with those of the modified webpage by calculating the \% change in \textit{AIR}.

\header{Results.} \Cref{tab:audit-change-results} highlights the percentage change in \textit{AIRs} after LLM modification of the webpages in our dataset. It presents the various audit categories (see~\Cref{sec:study_design}), the different models evaluated, and the percentage change in \textit{AIR} after modification. Negative percentages indicate successful issue resolutions and positive percentages suggest a regression of webpage performance. To easily identify performance, the worst regressions in each audit category are colored \textcolor{darkRed}{red}, and the best are \textcolor{darkGreen}{green}. Our findings are detailed below:

\begin{table*}
\centering
\caption{Percentage change in audit incidence ratio results (-ve values indicate reductions in incidents and represent improvements; +ve values indicate increases in incidents and represent regressions).}
\label{tab:audit-change-results}
\sisetup{
  round-mode            = places,
  round-precision       = 2,
  table-number-alignment= center
}
\begin{adjustbox}{max width=\linewidth}
\begin{tabular}{@{} l *{9}{S[table-format=-3.2]} @{}}
\toprule
\textbf{Audit Category}
  & {Claude 3.7 sonnet(R)}
  & {Claude 3.7 sonnet}
  & {Deepseek R1}
  & {Deepseek V3}
  & {GPT-4.1}
  & {GPT-4o-mini}
  & {Llama3.3 70B}
  & {o4-mini}
  & {Qwen2.5 32B-Instruct} \\
\midrule

SEO \& Accessibility               & \textcolor{darkGreen}{\textbf{-100.00}} & \textcolor{darkGreen}{\textbf{-100.00}} & \textcolor{darkGreen}{\textbf{-100.00}} & \textcolor{darkGreen}{\textbf{-100.00}} & \textcolor{darkGreen}{\textbf{-100.00}} & \textcolor{darkGreen}{\textbf{-100.00}} & \textcolor{darkGreen}{\textbf{-100.00}} & \textcolor{darkGreen}{\textbf{-100.00}} & \textcolor{darkGreen}{\textbf{-100.00}} \\
\midrule

Initial Load Performance               &  -18.57 &  -14.72 &  -23.97 &  -30.67 &  -46.52 &   \textcolor{darkRed}{24.84} &  -28.71 &  -26.22 &  \textcolor{darkGreen}{\textbf{-64.36}} \\
\midrule

Interactivity Performance               &   -5.00 &    6.09 &  -21.58 &  -21.09 &  -35.69 &    \textcolor{darkRed}{7.88} &  -54.69 &   -9.30 &  \textcolor{darkGreen}{\textbf{-64.99}} \\
\midrule

Runtime Performance               &  -11.11 &   -9.13 &  -32.62 &  -47.54 &  -37.38 &   \textcolor{darkRed}{58.97} &  -76.15 &  -31.83 &  \textcolor{darkGreen}{\textbf{-88.65}} \\
\midrule

Network Optimization               &  -14.35 &  -14.35 &  -33.40 &  -30.09 &  -48.68 &  -17.13 &  -35.05 &  -37.96 &  \textcolor{darkGreen}{\textbf{-64.68}} \\
\midrule

Resource Optimization               &    \textcolor{darkRed}{4.23} &   -8.47 &  -34.58 &  -11.33 &  -32.28 &   -4.94 &  -30.70 &  -20.50 &  \textcolor{darkGreen}{\textbf{-55.26}} \\
\midrule

Visual Stability               &   30.00 &   21.67 &   \textcolor{darkRed}{38.13} &   14.64 &  -22.02 &   28.33 &  -39.29 &   -6.03 &  \textcolor{darkGreen}{\textbf{-35.83}}\\
\bottomrule
\end{tabular}
\end{adjustbox}
\end{table*}

\textbf{Finding 1: LLMs universally excel at semantic understanding, resolving all SEO \& Accessibility issues.}
As shown in \Cref{tab:audit-change-results}, all LLMs achieved a \textbf{100.00\%} reduction for \textit{SEO \& Accessibility} audits. These types of issues typically do not rely on a comprehensive DOM context but rather on the semantic clarity and structural correctness of elements. This finding highlights the ability of LLMs to effectively understand and manipulate the semantic structure inherent within DOM elements, thus identifying and resolving issues crucial for SEO and web accessibility. This proficiency suggests that LLMs can be reliably employed to automate the correction of semantic markup, alt attributes for images~\cite{wu2017automatic, lopez2024turning}, ARIA roles, and other accessibility-related improvements without the need for exhaustive context. These capabilities are particularly valuable for maintaining compliance with accessibility standards and improving discoverability through search engines, aligning with prior research that emphasizes automated semantic validation and enhancement~\citep{wu2017automatic, wu2023ai}.

\textbf{Finding 2: High-performing LLMs (e.g., Qwen2.5 32B-Instruct, GPT-4.1) deliver significant, broad latency and optimization gains.}
Models such as \textbf{Qwen2.5 32B-Instruct} and \textbf{GPT-4.1} consistently demonstrated substantial improvements across multiple performance dimensions, positioning them as highly effective optimizers. These models, alongside \textbf{Claude 3.7 Sonnet(R)}, \textbf{Llama3.3 70B}, \textbf{Deepseek R1}, \textbf{Deepseek V3}, and \textbf{o4-mini} largely contributed to uniform decreases across all three latency audits, by substantial margins (Initial Load: from \textbf{-18.57\%} to \textbf{-64.36\%}; Interactivity: from \textbf{-5.00\%} to \textbf{-64.99\%}; Runtime: from \textbf{-11.11\%} to \textbf{-88.65\%}). 
While \textit{Runtime performance} saw the most significant individual reductions (up to \textbf{-88.65\%} for \textbf{Qwen2.5 32B-Instruct}), the variability in improvements was notably higher (standard deviation $\sigma=40.3$) compared to Initial Load ($\sigma=22.8$) and Interactivity ($\sigma=24.1$), indicating that even among improving models, the degree of enhancement varied.

Beyond latency, these high-performing LLMs also delivered significant gains in web asset delivery and transfer efficiency. Our analysis reveals a largely positive trend with consistent improvements in \emph{Network Optimization}. All evaluated LLMs demonstrated an ability to improve Network Optimization, with the reductions in \textit{AIR} ranging from \textbf{-14.35\%} (\textbf{Claude 3.7 Sonnet} and \textbf{Claude 3.7 Sonnet(R)}) to \textbf{-64.68\%} (\textbf{Qwen2.5 32B-Instruct}). For \emph{Resource Optimization}, most LLMs also delivered positive changes, with reductions ranging from \textbf{-4.94\%} (\textbf{GPT-4o-mini}) to \textbf{-55.26\%} (\textbf{Qwen2.5 32B-Instruct}). While \textbf{Claude 3.7 sonnet(R)} showed a slight increase of \textbf{+4.23\%}, the overall trend for the majority was positive. These findings underscore that LLMs largely possess the capability to enhance web asset delivery and transfer efficiency, demonstrating their advanced ability to generate DOM modifications that lead to more efficient asset delivery and consumption.

\textbf{Finding 3: GPT-4o-Mini presents a unique case of performance regression, particularly for user-facing latency and resource efficiency.}
In stark contrast to other LLMs, \textbf{GPT-4o-mini} consistently introduced significant overhead, leading to a notable \textit{regression} in crucial user-facing latency metrics and resource efficiency. Specifically, \textbf{GPT-4o-mini} increased the \textit{AIR} for all three user-facing speed audits: Initial Load (\textcolor{darkRed}{\textbf{+24.84\%}}), Interactivity (\textcolor{darkRed}{\textbf{+7.88\%}}), and dramatically for Runtime Performance (\textcolor{darkRed}{\textbf{+58.97\%}}). These increases represent the "Worst regression" cases for each of the audit categories. Manual inspection revealed these setbacks were primarily due to duplicated SVG path data that inflated payload size and paint cost, as well as the addition of new elements that bloat page sizes, indicating specific challenges this LLM faced in maintaining DOM integrity while optimizing.
Furthermore, \textbf{GPT-4o-mini} contributed to increased visual instability (\textcolor{darkRed}{\textbf{+28.33\%}}), reinforcing its tendency to introduce unintended DOM changes that degrade user experience. These outcomes highlight that while larger LLMs can translate performance optimization prompts into tangible latency savings, smaller LLMs may introduce new bottlenecks rather than eliminate existing ones. Accordingly, any production pipeline that relies on automated web issue resolution should pair model selection with post-hoc validation to prevent unintended speed/latency regressions.

\textbf{Finding 4: Visual stability remains a significant challenge for most LLMs, with a majority introducing regressions.}
The \emph{Visual Stability} audit captures unexpected layout shifts that harm perceived smoothness and can lead to frustrating user experiences. While some LLMs demonstrated proficiency in optimizing load times and network efficiency, the ability to maintain or improve \emph{Visual Stability} proved to be a pervasive challenge for the majority of LLMs. As shown in \Cref{tab:audit-change-results}, a clear trend emerged: most of the LLMs evaluated introduced some visual instability.
Most notably, \textbf{Deepseek R1} showed the most substantial regression in this category. Manual inspection revealed that many of these regressions stem from seemingly minor insertions or attribute changes that inadvertently alter element dimensions or flow. Common culprits include duplicated assets or scripts, changes in class names tied to cascading style sheet (CSS) styles, and the removal of scripts necessary for proper rendering. These issues highlight the inherent complexity of DOM manipulation and the difficulty LLMs currently face in consistently understanding spatial relationships within a webpage.

In contrast, four models achieved significant reductions in visual instability. The success of these models, particularly \textbf{Llama3.3 70B} with its nearly 40\% reduction, suggests that careful, targeted chunk-level modifications by certain LLM architectures may preserve or even enhance visual stability. However, their performance stands as an exception rather than the norm in this evaluation. These findings strongly underscore the critical need for robust post-processing checks whenever LLMs are employed for any form of DOM modification.
\vspace{-1.5ex}
\begin{tcolorbox}
\textbf{Answer to RQ1:} LLMs universally excel at semantic web issues, achieving a 100.00\% reduction in SEO \& Accessibility issues. For other audit categories, high-performing models (e.g., Qwen2.5 32B-Instruct, GPT-4.1) deliver significant gains, while notably, GPT-4o-mini consistently increased latency. The majority of LLMs introduced visual instability, underscoring the need for rigorous post-hoc validation for reliable web performance improvement.
\end{tcolorbox}

\section{RQ2: \rqii}
\label{sec:rqii}
\header{Motivation.} Building on RQ1's quantitative analysis of LLM effectiveness, RQ2 aims to understand the \textbf{nature of changes} LLMs make to the DOM. We saw LLMs achieve mixed results, with most improving performance while some introduced regressions ranging from increased latency to visual instability. This prompts a deeper dive into how LLMs modify the DOM and why specific outcomes occur. Investigating their DOM manipulations offers crucial insight into LLM's "black box," explaining effectiveness, revealing patterns, and informing future development, validation, and trust in automated web solutions.

\header{Approach.} To understand the nature of the LLM modifications, we parsed the DOM trees into structured JSON formats, we then generated detailed diffs between each original DOM tree and its modified version. This was done for the entire dataset. We used the open-source Python package \texttt{Deepdiff}, a robust tool for quantifying and classifying differences in hierarchical data like JSON DOM trees~\citep{deepdiff}. This systematic approach allowed us to identify and categorize the specific modifications introduced by each LLM, offering fine-grained insights into their interventions. The primary categories of changes defined by \texttt{Deepdiff} that we extracted for our analysis were:
\begin{itemize}[leftmargin=*]
    \item \textbf{\texttt{dictionary\_item\_added} / \texttt{dictionary\_item\_removed}:} These identify the addition or removal of \textbf{attributes} (e.g., \texttt{id}, \texttt{class}, \texttt{src}) on an HTML element.
    \item \textbf{\texttt{iterable\_item\_added} / \texttt{iterable\_item\_removed}:} These signify the addition or removal of \textbf{child elements or text nodes} within a parent element's content, often referred to as "element-level" or "node-level" changes.
    \item \textbf{\texttt{type\_changes}:} This category flags instances where the fundamental type of a DOM node (e.g., \texttt{<p>} node to a text node) was altered.
    \item \textbf{\texttt{values\_changed}} captures modifications to the content or properties of existing items that stay in place, broken down by type: \textbf{\texttt{attr\_changes}} (specific changes to existing attribute values, like \texttt{width="100"} to \texttt{width="50"}), \textbf{\texttt{tag\_changes}} (alterations to an element's HTML tag name, e.g., \texttt{<div>} to \texttt{<p>}), \textbf{\texttt{text\_changes}} (modifications to textual content within an HTML element or standalone text node), and the critical \textbf{\texttt{positional\_changes}} (quantifying reordering or shifts in item placement within a sequence due to additions or deletions).
\end{itemize}

\header{Results.}
For each LLM, we extracted the changes from all modified webpages, then grouped these changes by type, and finally summed the counts of each change type. We also report the depth of the changes performed by the LLMs across the dataset, as specified by \texttt{Deepdiff}. \Cref{tab:change-groups} highlights the categories of DOM changes; \textbf{Attributes} quantify \textbf{Added} (new attributes) and \textbf{Removed} (deleted attributes) on HTML elements; \textbf{Elements} summarize node-level modifications, covering \textbf{Added} and \textbf{Removed} elements; \textbf{Types Changed} indicate instances where a DOM node's fundamental type was altered; \textbf{Change Depth} reports the \textbf{Min}, \textbf{Max}, and median (\textbf{Med}) depth of change within the DOM tree for all modifications; and \textbf{Values Changed} break down modifications to existing DOM elements by specific type: \textbf{Attr} (attribute value changes), \textbf{Tag} (HTML tag name changes), \textbf{Pos} (positional changes), and \textbf{Text} (text content modifications). 

To quantify modification patterns, we introduce two metrics: the \textbf{Element Addition-to-Removal Ratio (EATRR)}, which indicates potential DOM bloat ($>0.5$) or simplification ($<0.5$) by comparing added versus removed elements; and \textbf{Positional Change Dominance (PCD)}, which measures the proportion of value changes attributed to reordering elements, highlighting disruption to spatial relationships. Our findings are detailed below:

\begin{table*}
\centering
\caption{Summary of Deepdiff Results by LLM Model with Calculated Ratios (including Depth stats)}
\label{tab:change-groups}
\small
\begin{adjustbox}{max width=\textwidth}
\begin{tabular}{
    l                            
    |S[table-format=1.2]         
    |S[table-format=0.2]         
    |*{2}{S[table-format=3]}     
    |*{2}{S[table-format=3]}     
    |S[table-format=1]           
    |*{3}{S[table-format=2]}     
    |*{4}{S[table-format=3]}     
    }
\hline
\textbf{Model}
  & {\textbf{EATRR}}
  & {\textbf{PCD}}
  & \multicolumn{2}{c|}{\textbf{Attributes}}
  & \multicolumn{2}{c|}{\textbf{Elements}}
  & \multicolumn{1}{c|}{\textbf{Types}}
  & \multicolumn{3}{c|}{\textbf{Change Depth}}
  & \multicolumn{4}{c}{\textbf{Values Changed}} \\
  & {}
  & {}
  & {Added} & {Removed}
  & {Added} & {Removed}
  & {Changed}
  & {Min} & {Max} & {Med}
  & {Attr} & {Tag} & {Pos} & {Text} \\
\hline
Claude 3.7 Sonnet(R)   & 0.89 & 0.42 &  2 &  1 &  90 &  11 & 4 &  1 & 12 & 4 & 10 &  7 &  57 &  24 \\
Claude 3.7 Sonnet      & 0.88 & 0.52 &  3 &  1 &  77 &  11 & 4 &  1 & 12 & 6 & 12 &  7 &  66 &  27 \\
Deepseek R1            & 0.50 & 0.31 &  4 &  7 &  37 &  37 & 7 &  1 & 37 & 4 & 13 &  8 &  37 &  32 \\
Deepseek V3            & 0.85 & 0.53 &  1 &  1 & 106 &  18 & 3 &  1 & 12 & 4 & 10 &  7 &  88 &  31 \\
GPT-4.1                & 0.89 & 0.48 &  2 &  1 & 111 &  14 & 4 &  1 & 12 & 4 & 10 &  8 &  79 &  31 \\
GPT-4o-mini            & 0.44 & 0.77 &  2 &  8 &  22 &  28 & 4 &  1 & 12 & 4 & 10 &  8 &  66 &  26 \\
Llama3.3 70B           & 0.67 & 0.40 &  3 &  3 &  90 &  45 & 4 &  1 & 24 & 6 & 10 &  8 &  74 &  41 \\
o4-mini                & 0.86 & 0.39 &  4 &  1 & 133 &  21 & 4 &  1 & 12 & 6 & 14 &  9 &  76 &  29 \\
Qwen2.5 32B-Instruct   & 0.74 & 0.47 &  2 &  1 &  70 &  25 & 4 &  1 & 24 & 6 & 10 &  8 &  70 &  45 \\
\hline
\end{tabular}
\end{adjustbox}
\end{table*}

\textbf{Finding 5: Most LLMs predominantly employ an additive DOM modification strategy.} As shown in the EATRR column of \Cref{tab:change-groups}, nearly all evaluated LLMs (except GPT-4o-mini and Deepseek R1) show a tendency to add significantly more elements than they remove. For instance, high-performing models like \textbf{Claude 3.7 Sonnet(R), Deepseek V3, GPT-4.1}, and \textbf{o4-mini} exhibit EATRRs ranging from $0.85$ to $0.89$. This implies that these models, in their pursuit of performance optimization, frequently introduce new elements, rather than primarily refactoring or simplifying the existing DOM. While \textbf{Deepseek R1} presents a more balanced approach with an EATRR of $0.50$, the overall inclination towards element addition suggests that these LLMs' optimization often involves enriching the DOM.

\textbf{Finding 6: GPT-4o-mini's exhibits a unique DOM modification strategy; removing more elements than it adds, coupled with the highest positional changes.} Among all evaluated models, \textbf{GPT-4o-mini} exhibits a distinctly unique DOM modification strategy. This strategy is characterized by the lowest EATRR at \textbf{0.44}, suggesting a tendency to remove or simply maintain existing element counts rather than an additive approach for optimization. Crucially, this is coupled with the highest PCD at \textbf{0.77}, indicating that a substantial majority of its modifications involve reordering or shifting existing elements. While other high-performing LLMs predominantly employ an additive strategy for optimization, GPT-4o-mini's distinct profile means it frequently attempts optimization through extensive and disruptive changes in element positioning. Such large-scale positional changes are known to be costly, often triggering expensive browser reflows and repaints, which directly contribute to visual instability and a poor Cumulative Layout Shift (CLS) score \citep{googleWebVitals, googlewebdom}. In contrast, models like Deepseek R1 (PCD 0.31, EATRR 0.50) and o4-mini (PCD 0.39, EATRR 0.89) exhibit different balances of positional changes and additive strategies. This unique combination of high, disruptive positional changes and a non-additive element strategy is a hallmark of GPT-4o-mini's behavior, which could explain its poor performance in Finding 3.

\textbf{Finding 7: LLMs operate across varying DOM depths, often concentrated at shallower levels.}
While LLMs make changes at various depths within the DOM tree, the median change depth across most models as shown in \Cref{tab:change-groups} ranges from 4-6, with minimum depths consistently at 1, and maximums ranging from 12 to 37. Depths over 32 are generally considered excessive for performance \citep{googlewebdom}. This indicates that LLMs are not just making superficial changes at the root level but are capable of intervening deeper within the DOM structure. However, they do not consistently reach the deepest possible levels. The variability in \texttt{Depth Max} (e.g., Deepseek R1 at 37 vs. GPT-4.1 at 12) suggests differences in how deeply models traverse and modify complex, nested structures. This overall pattern of intervention across various depths is a fundamental characteristic of LLM DOM manipulation.

\textbf{Finding 8: Textual modifications are a primary driver of performance and Visual Stability gains for effective LLMs.}
High-performing LLMs, such as \textbf{Qwen2.5 32B-Instruct} (PCD: $0.47$) and \textbf{GPT-4.1} (PCD: $0.48$), demonstrate that their success in improving performance metrics is strongly tied to the extent of their textual modifications. A very strong negative correlation exists between \texttt{Values Changed Text} and all three latency performance categories: \textbf{Initial Load} ($\rho = -0.74$), \textbf{Interactivity} ($\rho = -0.96$), and \textbf{Runtime} ($\rho = -0.97$), as well as \textbf{Network Optimization} ($\rho = -0.80$) and \textbf{Resource Optimization} ($\rho = -0.84$). This quantitatively suggests that LLMs making more textual changes are associated with greater performance improvements, indicating beneficial optimizations like minification of inline scripts or styles~\citep{googleWebVitals}. Furthermore, a strong negative correlation of $\rho = -0.90$ between \texttt{Values Changed Text} and Visual Stability for better-performing models in that audit category indicates these textual modifications also contribute to improved visual stability.

\begin{tcolorbox}
\textbf{Answer to RQ2:} Most of the LLMs evaluated primarily used an additive strategy for DOM modifications. The effective LLMs achieve performance gains via extensive textual modifications and at shallower DOM depths. In contrast, GPT-4o-mini shows a unique strategy, removing more elements than it adds, coupled with high positional changes. This is observed alongside its consistent performance regressions in RQ1.
\end{tcolorbox}


\section{Discussion}
\label{sec:discussion}
Our evaluation details the strengths and limitations of LLMs in automated web performance resolution, highlighting their effective applications and challenges for DOM manipulation. In the following, we discuss the key implications for research and practice.

\subsection{Proficiency in Semantic Understanding vs the Pervasive Challenge of Visual Stability}

The most consistent finding across all LLMs is their semantic understanding of the DOM (Finding 1). This proficiency underscores the LLMs' ability to grasp the semantic clarity and structural correctness of web elements, aligning with prior research on automated semantic validation and enhancement \citep{wu2017automatic, wu2023ai}. The implication here is profound; LLMs can be helpful when integrated into automated workflows for critical web development tasks that primarily involve semantic markup, alt attributes, ARIA roles, and other accessibility-related improvements, as well as for SEO considerations~\citep{delnevo2024interaction}. Some examples of related LLM changes we identified through manual inspection include adding \texttt{alt} descriptions to image elements~\citep{alahmadi2018evaluation} and introducing additional \texttt{meta} elements~\cite{roumeliotis2022effective}.

This can significantly reduce the manual effort and expertise required to maintain compliance with necessary web standards~\cite{w3cWCAG21, MDNMetaElement}. For developers and organizations, this translates into a powerful tool for proactive maintenance and adherence to best practices, potentially democratizing access to high-quality, accessible web content.

Despite their semantic prowess, a significant and widespread challenge identified in this study is the consistent struggle of most LLMs to maintain or improve \textbf{Visual Stability} (Finding 4). A majority of models introduced visual instability. Manual inspection revealed this largely stemmed from seemingly minor insertions or attribute changes that inadvertently altered element dimensions or flow, e.g., duplicated assets or scripts and prevalent changes in class names tied to CSS styles. This highlights a critical current limitation in LLMs' hierarchical and spatial understanding of the DOM. While they can semantically understand elements and generate code, they frequently fail to accurately predict the cascading visual effects of their modifications on page layout and rendering.

This limitation poses a substantial barrier to the full automation of web performance resolution, as visual stability is a critical component of performant websites~\citep{googleWebVitals}.
Future research must, therefore, intensively focus on enhancing LLMs' spatial reasoning and visual prediction capabilities within the context of DOM manipulation.

\subsection{Unpacking LLM Modification Strategies: Additive vs. Disruptive Approaches}

Our analysis of DOM modification types revealed distinct strategies employed by the LLMs in this study, is primarily additive.
Most LLMs predominantly employ an \textbf{additive DOM modification strategy} (Finding 5), introducing significantly more elements than they remove. While this approach can facilitate semantic enhancements or performance-critical additions, it carries the inherent risk of contributing to DOM bloat. Manual inspection revealed that a common cause of this bloat was the duplication of already existing SVG paths, which can counteract performance gains in other areas. Furthermore, while LLMs operate across varying DOM depths (Finding 7), some models like Deepseek R1 showed extreme maximum depths of change (up to 37), exceeding the advised maximum depth of 32 to mitigate increased memory usage caused by large DOMs~\citep{googlewebdom}. This indicates that while LLMs are capable of deep interventions, such extreme depth changes, even if related to positional shifts, warrant caution due to potential performance overhead.

In contrast to this additive trend, \textbf{GPT-4o-mini exhibited a uniquely disruptive strategy}, characterized by removing more elements than it adds, coupled with the highest PCD (Finding 6). This frequent reordering and shifting of elements, without a corresponding reduction in overall element count, is a likely contributor to its consistent performance regressions observed in RQ1 (Finding 3), especially for user-facing latency and visual stability. Such large-scale positional changes are known to be computationally expensive, often triggering costly browser reflows and repaints, which directly degrade overall user experience~\citep{googlewebdom}.

Conversely, the success of high-performing LLMs like \textit{Qwen2.5 32B-Instruct} and \textit{GPT-4.1} in improving performance metrics is strongly tied to their extensive textual modifications (Finding 8). These models excel when optimizations can be expressed primarily through textual manipulation, often at shallower, more impactful DOM levels. Manual inspection confirmed this, showing these models adding performance-critical attributes (e.g., \texttt{defer} and \texttt{async} for faster script loading times) to \texttt{link}, and \texttt{script} elements often found in the \texttt{head} section of webpages, closer to the DOM root. This strategic placement and modification of existing attributes directly contribute to improved load and runtime performance~\citep{googlewebdom}.

\subsection{Overall Implications for Automated Web Performance Resolution}

The findings present LLMs as powerful yet incomplete tools for automated web performance remediation. They excel at \textit{semantic} optimizations (e.g., SEO, accessibility) but remain unreliable for \textit{performance-critical} tasks (e.g., load speed, interactivity), where regressions—especially in visual stability—were frequent. These results highlight the need for careful integration rather than full automation. The key implications are:
\begin{itemize}[leftmargin=*]
    \item \textbf{Hybrid Approaches are Essential:} Given the mixed outcomes, LLM-driven optimization should be paired with automated validation (e.g., Lighthouse) and, when needed, human oversight. CI/CD pipelines must include checks for visual stability and latency to detect and prevent regressions introduced by LLM edits.
    \item \textbf{Targeted LLM Deployment:} LLMs are best applied to well-defined tasks where their strengths are proven, such as semantic optimization. For more sensitive performance areas, careful model selection and rigorous testing are crucial. As shown by differences between high-performing models and \textit{GPT-4o-Mini}, characterizing a model’s typical DOM modification patterns (e.g., EATRR, PCD, depth of edits) can guide safer deployment.
    \item \textbf{Need for Enhanced LLM Capabilities:} Future work should address limitations in hierarchical and spatial reasoning to improve visual stability. Promising directions include architectural changes for better layout awareness and prompt strategies that explicitly account for visual impact.
\end{itemize}

In conclusion, LLMs hold strong potential to accelerate web optimization but require thoughtful, guarded integration into existing workflows and continued research to close current capability gaps.

\section{Limitations}
\label{sec:limitations}

\header{Internal Validity.} Threats to internal validity concern factors that may affect the accuracy of our findings. A key source of bias is the sensitivity of LLM performance to prompt design. We mitigated this by using consistent, well-tested zero-shot prompts across all models (\Cref{sec:study_design}). Temperature was fixed at 0.0~\citep{doderlein2025piloting}, and single trials were used for cost reasons, which may understate variance. Output and chunk limits matched the smallest context window (GPT-4o-mini) to ensure fairness; future work should explore larger contexts and repeated runs. To prevent context loss when handling large DOM trees, we applied a structured chunking strategy validated through Tree Edit Distance checks~\citep{shin2021learning, leithner2018domdiff}.

\header{External Validity.} Threats to external validity concern the generalizability of our findings. These depend on the dataset, LLMs, and audit tools used. We evaluated 15 popular, real-world webpages~\citep{alexawebsites} that varied widely in use, structure, and complexity, and nine state-of-the-art LLMs differing in size, context limits, and reasoning ability. Although our results represent a snapshot of current model versions, this breadth supports meaningful generalization. To ensure consistency, we used Lighthouse~\citep{lighthouseOverview} as the sole auditing tool, enabling comparability with prior studies~\citep{indonesiaLighthouse, towardsImprovingAccessibility, saif2021early}. Our study did not measure direct user experience metrics or production factors such as computational cost or inference latency~\citep{albert2022measuring, wehner2022vital}, which remain promising directions for future work.

\section{Related Work}
\label{sec:related_work}

Recent advancements in LLMs have significantly expanded their applicability to web development, moving beyond traditional natural language processing to encompass a deeper understanding and manipulation of the DOM~\citep{hou2023large, seAdapt2024, hadi2024large}. We group the related work into three major themes: LLMs for HTML Understanding, LLMs for Web Content Generation and Customization, and LLMs for Web-Based Task Automation and Information Extraction.

\subsection{LLMs for HTML Understanding}
LLMs are increasingly used for HTML understanding, parsing raw HTML for tasks like web-based automation and browser-assisted retrieval~\citep{Gur2022UnderstandingHW, ahluwalia2024leveraging, nass2023improving}. This capability hinges on an LLM's understanding of HTML's semantic structure, tag-based syntax, and hierarchical organization (forming the DOM tree).
Notably, \citet{Gur2022UnderstandingHW} showed that LLMs pre-trained on natural language can readily transfer to HTML understanding, requiring minimal preprocessing for tasks like semantic classification and description generation.
Our study builds on this by assessing how LLMs, using raw HTML, not only comprehend its structure and semantics but also apply this understanding to resolve web performance issues, requiring a deeper, actionable interpretation of HTML elements in context.

\subsection{LLMs for Web Content Generation and Customization}
LLMs have been leveraged to generate and customize web content. \citet{calo2023leveraging} demonstrated LLMs' ability to facilitate website creation from natural language descriptions, validating their HTML understanding for this task. Similarly, \citet{customizeHTML2023} explored LLMs for UI customization, enabling style-related DOM changes through natural language commands.
Our research goes a step further by focusing on web performance optimization. We evaluate how well LLMs can modify the DOM to ensure faster style changes for subsequent UI customizations.

\subsection{LLMs for Web-Based Task Automation and Information Extraction}
LLMs have shown promise in web automation and information extraction. \citet{nass2023improving} showed LLMs can enhance web element identification in GUI test automation by localizing elements based on contextual awareness. Research on information extraction, such as WebFormer by \citet{wang2022webformerMichel}, often focuses on text content but also highlights the importance of leveraging HTML's structural and layout information for accuracy.
Our study expands these applications by evaluating LLMs' capacity to make actionable DOM modifications to resolve complex, performance-related issues.

\section{Conclusion}
\label{sec:conclusion}

We evaluated nine state-of-the-art LLMs for automated resolution of web performance issues. The models demonstrated universal proficiency in SEO and accessibility optimization, leveraging strong semantic and structural DOM understanding. However, their effectiveness on performance-critical issues (latency, network, resource) was highly variable; most introduced visual instability due to limited spatial reasoning. We observed predominantly additive modification strategies—sometimes leading to bloat—with \textit{GPT-4o-mini} exhibiting uniquely disruptive, high-positional-change behavior. Effective optimizations arose from models capable of targeted, depth-aware edits, underscoring the need for precision and post-hoc validation before deployment. We acknowledge that web performance depends on the interaction of DOM, CSS, and JavaScript. Our DOM-centric scope was a deliberate choice to provide a controlled, reproducible benchmark \textit{without confounding build or server-side factors}. This focus established a foundational baseline for isolating LLM behaviors, while the observed regressions reveal the challenges of automated DOM manipulation. Extending future work to richer CSS/JS contexts is a natural next step toward a more complete understanding of LLM-driven web optimization.

\bibliographystyle{ACM-Reference-Format}
\bibliography{references}

\onecolumn
\appendix





\end{document}